\documentclass{webofc}
\usepackage[varg]{txfonts}   
\usepackage{comment}
\usepackage{color}
\usepackage{wrapfig}

\begin{document}
\title{New theoretical developments on the early-time dynamics and approach to equilibrium in Heavy-Ion collisions}
%
%

\author{
        \firstname{S\"{o}ren} \lastname{Schlichting }\thanks{E-Mail:sschlichting@physik.uni-bielefeld.de}
}

\institute{
           Fakult\"{a}t f\"{u}r  Physik, Universit\"{a}t Bielefeld, D-33615 Bielefeld, Germany
          }

\abstract{%
We discuss recent theoretical developments in understanding the early pre-equilibrium dynamics and onset of hydrodynamic behavior in high-energy heavy-ion collisions. We highlight possible experimental signatures of the pre-equilibrium phase, and present recent progress in developing a consistent theoretical description of collective flow in small systems.}
\maketitle
\section{Introduction}
Since the dynamical description of heavy-ion collisions from the underlying theory of Quantum Chromo Dynamics (QCD) remains an outstanding challenge, the standard model of nucleus-nucleus (A+A) collisions~\cite{Heinz:2013wva} is based on a hierarchy of effective descriptions of the underlying QCD dynamics, which exploits a clear separation in the time scale of the reaction dynamics. Before the collision, the atomic nuclei are described 
in terms of their partonic content of quarks and gluons, which interact with each other over the course of $\tau_{\rm coll} \ll 1$ of the actual collision of the nuclei, depositing a large amount of energy into the initial state of the subsequent evolution. Subsequently, the highly excited non-equilibrium QCD state undergoes a pre-equilibrium evolution, before on time scales $\tau_{\rm Hydro} \sim 1 {\rm fm}/c$ a near-equilibrium Quark-Gluon Plasma (QGP) is formed, whose subsequent space-time evolution can be described in terms of relativistic viscous hydrodynamics. Eventually, the QGP cools by long. and transverse expansion, before on time scales $\sim 10 {\rm fm}/c$, it undergoes hadronization and shortly after freezes out, to produce the observed particle spectra.

Based on the above picture of the reaction dynamics, it is clear that a dynamical description of the pre-equilibrium stage, is highly desirable to enable a continuous description that connects the cold QCD properties of nuclei with the hot QCD matter studied primarily in the collisions. 

\section{Equilibration in Heavy-Ion collisions — Kinetic \& Chemical Equilibrium}
The non-equilibrium QCD plasma created almost immediately after the collision of heavy nuclei, is believed to be kinetically and chemically out of equilibrium.  Due to a preference for radiative gluon emission, the initial state is expected to be highly gluon dominated, whereas in thermal equilibrium quarks degrees of freedom give the dominant contribution to the thermodynamic properties of the QGP.  Due to the rapid longitudinal expansion, the pre-equilibrium QGP is initially highly anisotropic and unable to sustain a sizeable longitudinal pressure $P_L$.  By neglecting the transverse expansion of the system over the short $\sim 1{\rm fm}/c$ period of the pre-equilibrium evolution, there has been significant progress in recent years in understanding the approach of the pre-equilibrium QGP towards kinetic and chemical equilibrium.

\begin{wrapfigure}[20]{r}{0.5\textwidth}
\centering\includegraphics[width=0.5\textwidth]{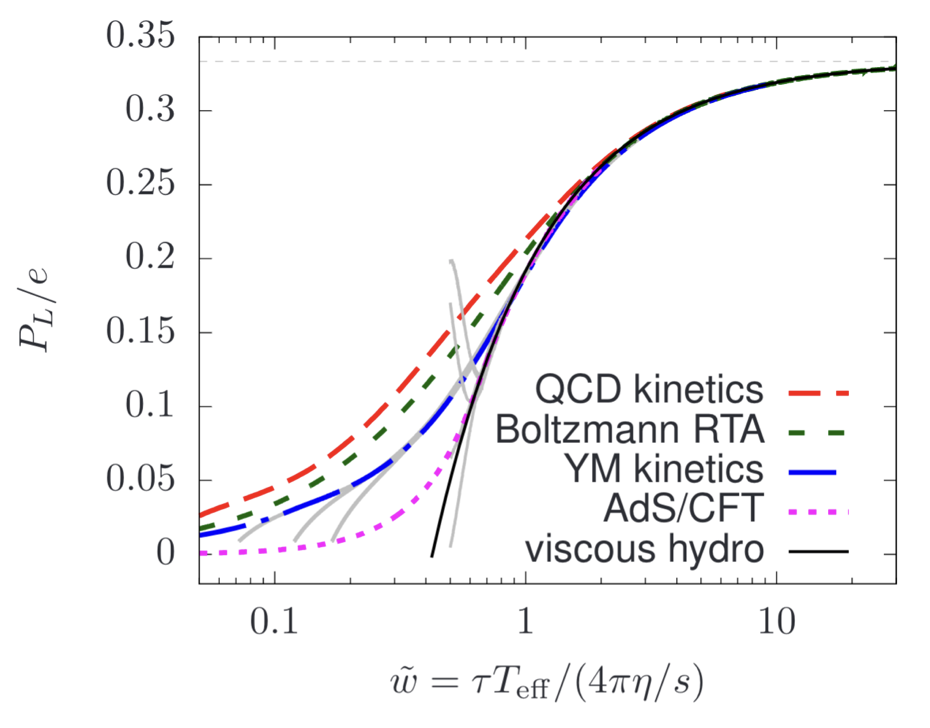}
    \caption{Evolution of the long. pressure $P_L$ to energy $e$ density ratio, during the pre-equilibrium phase. Different microscopic theories show the onset of visc. hydrodynamic behavior on time scales $\sim \frac{4\pi \eta/s}{T}$. ~From \cite{Giacalone:2019ldn}}
    \label{fig:Attractor}
\end{wrapfigure}
\textit{Kinetic equilibration of the QGP} While initially the long. pressure $P_L$ is much smaller than the energy density $\epsilon$, it is build up dynamically due to interactions over the course of the pre-equilibrium stage. Different calculations performed at weak and strong coupling now provide a consistent theoretical picture of how this approach towards equilibrium occurs on a macroscopic level. In order to meaningful compare the dynamics at different coupling strength, it is instructive to consider the evolution in terms of the conformal scaling variable $\tilde{w}=T(\tau)\tau/4\pi \eta/s$, which can be seen as the evolution time $\tau$ measured in units of the equilibrium relaxation time $\tau_{R}^{\rm eq}=4\pi \eta/s / T$. While at early times $\tilde{w} \ll 1$ the pre-equilibrium QGP is dominated by the longitudinal expansion, and behaves as macroscopically free-streaming with $P_L/\epsilon \approx 0$, towards later times $\tilde{w} \gtrsim 1$ the QGP becomes describable by visc. relativistic hydrodynamics, where (for a conformal system) $P_L/e = 1/3 - 4/9\pi\tilde{w}$. Different microscopic calculations interpolate slightly differently between these two limiting behaviours, as can be seen in Fig.~\ref{fig:Attractor}, where the evolution of $P_L/\epsilon$ is compared for Yang-Mills (YM)~\cite{Kurkela:2015qoa,Kurkela:2018vqr}  and QCD kinetic theory~\cite{Kurkela:2018xxd,Du:2020zqg}, the (conformal) relaxation time approximation (RTA)~\cite{Strickland:2018ayk,Kamata:2020mka} and strongly coupled $\mathcal{N}=4$ SYM theory (AdS/CFT)~\cite{Romatschke:2017vte,Beuf:2009cx}. Notably, in all of these cases, different initial conditions converge to the same curve for a given microscopic theory, giving rise to a so called ``hydrodynamic attractors’’~\cite{Heller:2015dha,Kurkela:2019set,Soloviev:2021lhs}, which characterises the fact that there is a rapid memory loss of certain aspects of the initial conditions, such as the initial pressure anisotropy, that occurs prior to the genuine onset of equilibration. 

Evidently, the most important information obtained from these studies, is the fact that viscous hydrodynamics is able to describe the evolution of the system on time scales $\tilde{w} \sim 1$. Stated differently, these studies indicate that hydrodynamics becomes applicable on time scales of the order of a single equilibrium relaxation time $\tau_{R}^{\rm eq}$, which in terms of phenomenological estimates for LHC energies translates to 
$$
\tau_{\rm Hydro} \approx 1.1 {\rm fm}/c \left( \frac{4\pi eta/s}{2} \right)^{3/2} \left( \frac{(s\tau)_{\rm \infty}}{4.1 {\rm GeV}^2} \right)^{-1/2}
$$
where $\eta/s$ is the shear-viscosity to entropy density ratio, and $(s\tau)_{\rm \infty}$ is the (asymptotic) entropy per unit rapidity. However, it should also be noted, that even at these times, the QGP is still significantly out-of-equilibrium, and features a residual order one anisotropy. 

New theoretical developments in the approach to equilibrium also include a first proof-of-principle calculation of the evolution of the pressure anisotropy in scalar quantum field theory~\cite{Gelis:2023bxw}, as well as a detailed characterisation of attractors in QCD kinetic theory~\cite{Boguslavski:2023jvg}, 
which indicates the emergence of (slightly) different attractors in the weak/strong coupling limit, reflecting the presence/absence of a separation of scales. While so far, most theoretical studies have focused on kinetic equilibration for conformal systems, new theoretical works ~\cite{Alalawi:2022pmg,Jaiswal:2022udf} have also demonstrated the persistence of hydrodynamic attractors for the relaxation time approximation with massive quasi-particles, and it would be interesting if challenging to extend these studies towards more realistic QCD interactions.

\textit{Chemical equilibration of the QGP} By now different works have also studied the chemical equilibration of the QGP~\cite{Kurkela:2018xxd,Du:2020zqg} and found that the latter occurs essentially on the same time scale $\tilde{w} \sim 1-1.5$ as hydrodynamics becomes applicable. However, since interactions are required to change the chemical composition of the pre-equilibrium QGP, an initial gluon dominance persist up to times $\tau_{\rm hydro}$. Beyond these phenomenologically relevant findings, there are also detailed microscopic studies of chemical equilibration in QCD kinetic theory, and we refer the interested reader to~\cite{Du:2020dvp,Cabodevila:2023htm}.

\textit{Dynamical description of the pre-equilibrium stage — KoMPoST} Based on the theoretical progress outlined above, the early pre-equilibrium stage can also be included in event-by-event simulations of heavy-ion collisions, using the effective macroscopic description of KoMPoST~\cite{Kurkela:2018wud,Kurkela:2018vqr}. Central to this framework is the idea, that it is sufficient to study the macroscopic evolution of the energy-momentum tensor, which can be evolved using non-equilibrium linear response theory. By decomposing the initial energy momentum tensor $T^{\mu\nu}$ into a locally homogeneous and boost invariant average $\bar{T}^{\mu\nu}$, and linearized fluctuations $\delta T^{\mu\nu}$, the energy-momentum tensor that serves as initial condition for the subsequent hydrodynamic evolution can then be obtained as
\begin{eqnarray}
\label{eq:TMunNu}
    T^{\mu\nu}_{\rm Hydro}(\tau,{\bf x})= \bar{T}^{\mu\nu}\big({\bf x},\tilde{w}(\tau,{\bf x})\big) + \int_{\odot} d^2{\bf x}_o G^{\mu\nu}_{\alpha\beta}\big(\tau-\tau_0,{\bf x}-{\bf x}_0,\tilde{w}(\tau,{\bf x})\big)\delta T^{\alpha\beta}({\bf x},{\bf x}_0)
\end{eqnarray}
where $G^{\mu\nu}_{\alpha\beta}$ are non-equilibrium Green's functions calculated in kinetic theory. While the original calculation of KoMPoST provided a calculation of these Green's functions in pure glue QCD kinetic theory, these calculations have recently been updated to full QCD kinetic theory~\cite{Dore:2023qxr}. Strikingly, calculations in kinetic theory in the relaxation time approximation (RTA), also show essentially the same behavior for the non-equilibrium Green's functions~\cite{Kamata:2020mka}, which was further explored in ~\cite{Du:2023bwi} where a universal behavior of equilibrium Green's functions of the energy momentum tensor in kinetic theory was demonstrated. Since kinetic theory in the relaxation time approximation (RTA) is sufficiently can also be solved non-linearly for non-trivial spatial profiles, the KoMPoST framework was further benchmarked again fully microscopic calculations in~\cite{Ambrus:2022koq}, yielding an excellent agreement for almost all components of the energy-momentum tensor. 

Specifically, for perturbations of the energy-momentum tensor around thermal equilibrium, it is also possible to decompose the response of the system into hydrodynamic and non-hydrodynamics modes~\cite{Kurkela:2017xis,Romatschke:2015gic} and thereby assess the range of applicability of effective macroscopic descriptions. Despite the fact that the spectrum of non-hydrodynamic excitation in kinetic theory turns out to be rather complicated~\cite{Kurkela:2017xis,Ochsenfeld:2023wxz}, it turns out that the corresponding Green's functions in the sound channel show a remarkable degree of universality between different microscopic theories and can effectively be described by a single non-hydrodynamic mode in addition to the hydrodynamic mode~\cite{Du:2023bwi}. Nevertheless, the non-hydrodynamic response dominates below size scales $\ell \lesssim 0.16 {\rm fm} \left(\frac{200 {\rm MeV}}{T} \right)  \left(\frac{\eta/s}{0.16} \right)$, indicating that hydrodynamics is no longer applicable on such small scales~\cite{Ochsenfeld:2023wxz}.

One important phenomenological development concerns the inclusion of conserved charges into the pre-equilibrium evolution o heavy-ion collisions. By expandind the charge currents $J^{\mu}$ around a charge neutral background $(\bar{J}^{\mu}=0)$, the initial conditions for the hydrodynamic evolution can be obtained in analogy to Eq.~(\ref{eq:TMunNu}) as 
\begin{eqnarray}
    J^{\mu}_{\rm Hydro}(\tau,{\bf x})= \int_{\odot} d^2{\bf x}_o~F^{\mu}_{\alpha}\big(\tau-\tau_0,{\bf x}-{\bf x}_0,\tilde{w}(\tau,{\bf x})\big)\delta J^{\alpha\beta}({\bf x},{\bf x}_0)
\end{eqnarray}
which was recently incorporated into KoMPoST~\cite{Dore:2023qxr}. Evidently, with including conserved charges, there is also a need to implement the longitudinal response and extent the KoMPoST framework to 3+1D. However, this question has not been addressed to date.

\section{Phenomenology of the pre-equilibrium phase in HICs}
Based on the theoretical progress in understanding different aspects of the early-time pre-equilibrium dynamics in heavy-ion collisions, recent works have started to develop a phenomenology of the pre-equilibrium phase in heavy-ion collisions. Even though, the prospect of investigating non-equilibrium QCD processes in heavy-ion collision experiments is certainly appealing, this is complicated by the fact that the short pre-equilibrium phase of $\sim 1 {\rm fm}/c$ at the beginning of the collision typically has a very small impact on bulk observables, which are dominated by the hydrodynamic expansion of the QGP and hadronic re-scattering~\cite{Kurkela:2018vqr}. By considering very specific observables, it is nevertheless possible to augment the role of the pre-equilibrium phase as illustrated at the examples below.

\emph{Entropy production}

\emph{Electromagnetic probes} Electromagnetic probes, such as photons and dileptons, escape the fireball unscathed, are therefore sensitive to all stages of the space-time evolution. Despite the fact that electromagnetic probes are produced throughout the entire history of the collision, it is possible to augment the contribution of the pre-equilibrium phase by focusing on particular kinematics. Specifically, recent pheonomenological calculations for photons~\cite{Garcia-Montero:2023lrd} show a sizeable pre-equilibrium contribution for transverse momenta $3 {\rm GeV} \lesssim p_T \lesssim 6 {\rm Gev}$ at LHC energies, which can be comparable to the prompt contribution from initial hard scattering. Exploiting a universal scaling behavior of the pre-equilibrium contributions, it is now possible to perform such calculations on an event-by-event basis, and the corresponding results have been incorporated into (Shiny)KoMPoST.

\begin{wrapfigure}[20]{r}{0.5\textwidth}
\centering\includegraphics[width=0.5\textwidth]{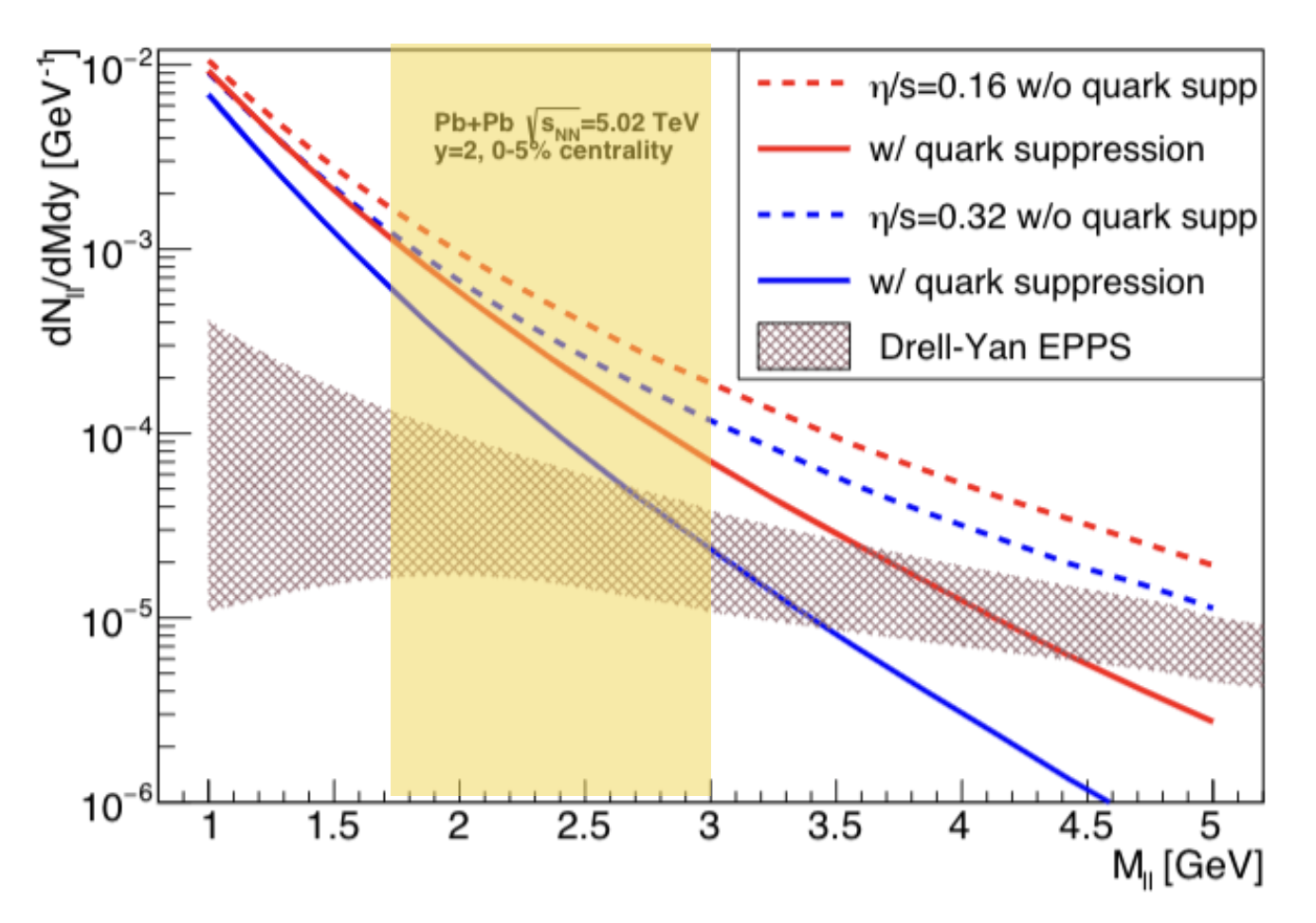}
    \caption{Dilepton yield $dN_{ll}/dydM$ as a function of invariant mass $M_{ll}$ for Pb+Pb collisions at LHC energies. Highlighted is the invariant mass range, where pre-equilibrium production dominates over Drell-Yan and thermal contributions.~From \cite{Coquet:2021lca}}
    \label{fig:Dilepton}
\end{wrapfigure}
Nevertheless, intermediate mass dileptons remain the most promising candidate for a direct detection of signals from the pre-equilibrium calculations, as recent phenomenological calculations~\cite{Coquet:2021lca} shown in Fig.~\ref{fig:Dilepton} suggest a dominant contribution from the pre-equilibrium phase for invariant masses $1.5 {\rm GeV}  \lesssim  M \lesssim 3 {\rm GeV}$. Strikingly, a more differential investigation of dileptons considering the so-called polarization in the Collins-Soper frame~\cite{Coquet:2023wjk} for this invariant mass range, can further help to distinguish between different production mechanisms, and may even provide a direct measure of the momentum anisotropy of the pre-equilibrium QGP. Even though, the measurement of dileptons in this invariant mass range is subject to a large background from heavy-flavor decays, it appears to be likely that such measurements will become feasible with prospective LHC detector upgrades~\cite{LHCb:2018roe,ALICE:2022wwr}
in the next decade. 

\textit{Hard probes} Beyond the aforementioned aspects, theoretical studies have also started to explore the effect of the pre-equilibrium on heavy-flavors and jets and in both cases found sizeable and anisotropic contributions to momentum broadening~\cite{Boguslavski:2023fdm,Boguslavski:2023alu}. However, so far there is no genuine signature that distinguishes the contributions of the pre-equilibrium phase from the subsequent evolution.

\section{Equilibration in Small systems? — Non-equilibrium descriptions of coll. flow \& (In)applicability of hydrodynamics}

Due to the smaller lifetime of the system, which is determined by the (transvserse) system size $R$, the clear separation of time scales in the reaction dynamics ceases to exist in smaller collision systems, and it is thus not clear to what extent the standard picture of the space-time evolution of heavy-ion collisons~\cite{Heinz:2013wva}, is applicable to e.g. $p+p$ or $p+Pb$ collisions. Since it is a legitimate possibility that the system may fall apart prior to undergoing a sufficient degree of equilibration,  this prompts the development of more microscopic treatments of the collective expansion of the out-of-equilibrium  QGP. Evidently, it is challenging to address the 3+1D space-time dynamics microscopically but
first calculations in the weak and strong coupling limit have been pioneered already~\cite{Chesler:2015bba,Chesler:2015wra,Greif:2017bnr,Kurkela:2021ctp,Tornkvist:2023kan}. While this question has also prompted the development of a new QCD Kinetic Theory Monte-Carlos~\cite{Tornkvist:2023ylh}, a recent systematic study in kinetic theory in the (conformal) relaxation time approximation provides important insights on the (in)applicability of hydrodynamics in small systems~\cite{Ambrus:2022qya,Ambrus:2022koq}. Strikingly, in this simplistic microscopic description, the 
collective dynamics of the system only depends on the initial spatial geometry, and a single dimensionless opacity parameter $\hat{\gamma} \propto (\eta/s)^{-1} R^{1/4} ({dE^{0}_{\bot}}/{d\eta})^{1/4}$ which measures the interaction strength, and encodes the dependence on the shear-viscosity to entropy density $\eta/s$, system size $R$ and initial energy per unit rapidity $dE^{0}_{\bot}/{d\eta}$~\cite{Ambrus:2022koq,Ambrus:2021fej,Kurkela:2018qeb,Kurkela:2019kip}.

\begin{wrapfigure}[21]{r}{0.5\textwidth}
    \centering
    \includegraphics[width=0.5\textwidth]{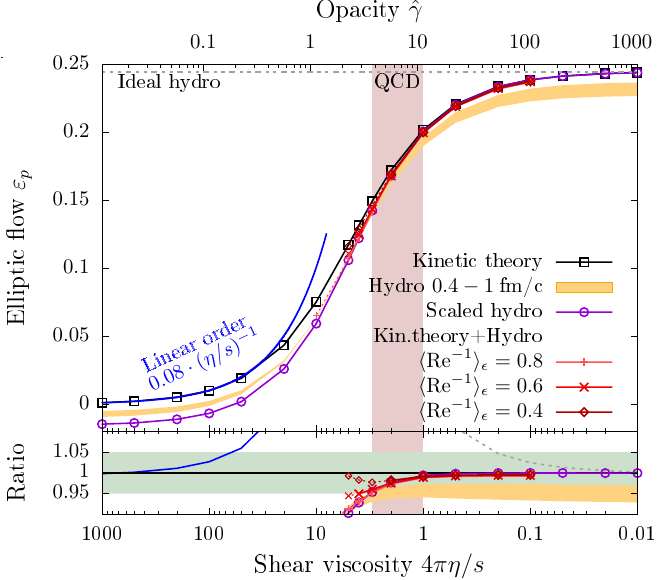}
    \caption{Development of collective flow in kinetic theory (RTA) as a function of opacity $\hat{\gamma}$. Beyond opacities $\hat{\gamma}\gtrsim 3-4$, the collective flow can be quantitatively described by matching kinetic theory (at early times) to hydrodynamics (at late times). From\cite{Ambrus:2022qya}}
    \label{fig:CollFlow}
\end{wrapfigure}

By varying the opacity over several orders of magnitude from the non-interacting limit $(\hat{\gamma} \to 0)$, towards ideal hydrodynamics $(\hat{\gamma} \to \infty)$, one observes a smooth onset of collective flow, as seen in Fig.~\ref{fig:CollFlow}. Clearly, this demonstrated that the onset of collective behavior is not indicative of  hydrodynamic expansion of the QGP, but instead in becomes a quantitative question, to what extent the underlying microscopic dynamics can be accurately described by an effective hydrodynamic description. By comparing of  the results fully microscopic calculations to hybrid simulations, matching kinetic theory to hydrodynamics at different values of the (average) inverse Reynolds number $Re^{-1}=\sqrt{\frac{6\pi^{\mu\nu}\pi_{\mu\nu}}{e^2}}$, also shown in Fig.~\ref{fig:CollFlow}, one concludes that opacities $\hat{\gamma} \gtrsim 3-4$ are necessary for hydrodynamics to provide a meaningful and quantitatively accurate description of collective flow. By comparing these results to the typical opacities reached in $p+p$ or $p+Pb$ collisions, one concludes that hydrodynamics is not applicable in such small systems~\cite{Ambrus:2022qya}. Interestingly, oxygen-oxygen (O+O) collision are expected to fall into the transition region, and will thus provide an interesting opportunity to study onset of hydrodynamic behavior.

\section{Conclusions \& Outlook}
Significant theoretical progress has been achieved in the theoretical description of the early pre-equilibrium phase of heavy-ion collisions, which now enables the inclusion of a pre-equilibrium phase into state-of-the-art theoretical descriptions, and gives rise to an interesting phenomenology to be explored with present and future heavy-ion experiments. However, the studies also indicate that hydrodynamics is not applicable in small systems, as the QGP does not equilibrate to sufficient degree, thus challenging the community to develop an adequate theoretical description to describe the non-equilibrum QCD dynamics of small systems.\\

\textit{Acknowledgements:} We acknowledge support by the Deutsche Forschungsgemeinschaft (DFG, German Research Foundation) through the CRC-TR 211 ‘Strong-interaction matter under extreme conditions’-project number 315477589 – TRR 211 and by the German Bundesministerium für Bildung und Forschung (BMBF) through Grant No. 05P21PBCAA

\bibliography{template}
\end{document}